\pdfoutput=1
\documentclass[12pt,a4paper]{article}
\usepackage{ifthen} 
\usepackage{float}
\newboolean{pdflatex}
\setboolean{pdflatex}{true} 

\newboolean{articletitles}
\setboolean{articletitles}{true} 

\newboolean{uprightparticles}
\setboolean{uprightparticles}{false} 

\newcommand{\unitm}[1]{\,{\rm{#1}}}

\newcommand{\bppk}{B^{+} \to p \bar p K^{+}}
\newcommand{\mpp}{M_{p \bar p} < 2.85}

\usepackage{microtype}
\usepackage{lineno}  
\usepackage{xspace} 

\usepackage{graphicx}  
\usepackage{color}
\usepackage{colortbl}
\graphicspath{{./figs/}} 

\usepackage{amsmath} 
\usepackage{amssymb}
\usepackage{amsfonts}
\usepackage{upgreek} 

\newcommand*\patchAmsMathEnvironmentForLineno[1]{%
\expandafter\let\csname old#1\expandafter\endcsname\csname #1\endcsname
\expandafter\let\csname oldend#1\expandafter\endcsname\csname
end#1\endcsname
 \renewenvironment{#1}%
   {\linenomath\csname old#1\endcsname}%
   {\csname oldend#1\endcsname\endlinenomath}%
}
\newcommand*\patchBothAmsMathEnvironmentsForLineno[1]{%
  \patchAmsMathEnvironmentForLineno{#1}%
  \patchAmsMathEnvironmentForLineno{#1*}%
}
\AtBeginDocument{%
\patchBothAmsMathEnvironmentsForLineno{equation}%
\patchBothAmsMathEnvironmentsForLineno{align}%
\patchBothAmsMathEnvironmentsForLineno{flalign}%
\patchBothAmsMathEnvironmentsForLineno{alignat}%
\patchBothAmsMathEnvironmentsForLineno{gather}%
\patchBothAmsMathEnvironmentsForLineno{multline}%
}

\usepackage{hyperref}    
\usepackage[all]{hypcap} 

\def\lhcb {\mbox{LHCb}\xspace}
\def\ux85 {\mbox{UX85}\xspace}

\ifthenelse{\boolean{uprightparticles}}%
{

 \def\Peta        {\ensuremath{\upeta}\xspace}

 \def\Ppsi        {\ensuremath{\uppsi}\xspace}

 \def\PDelta      {\ensuremath{\Delta}\xspace}                 
 \def\PXi      {\ensuremath{\Xi}\xspace}                 
 \def\PLambda      {\ensuremath{\Lambda}\xspace}                 
 \def\PSigma      {\ensuremath{\Sigma}\xspace}                 
 \def\POmega      {\ensuremath{\Omega}\xspace}                 
 \def\PUpsilon      {\ensuremath{\Upsilon}\xspace}                 
 
 \def\PB      {\ensuremath{\mathrm{B}}\xspace}                 
                  
 \def\PD      {\ensuremath{\mathrm{D}}\xspace}

 \def\PJ      {\ensuremath{\mathrm{J}}\xspace}                 
 \def\PK      {\ensuremath{\mathrm{K}}\xspace}

 \def\Pb      {\ensuremath{\mathrm{b}}\xspace}                 
 \def\Pc      {\ensuremath{\mathrm{c}}\xspace}

 \def\Pi      {\ensuremath{\mathrm{i}}\xspace}

}
{

 \def\Peta        {\ensuremath{\eta}\xspace}

 \def\Ppsi        {\ensuremath{\psi}\xspace}                 
                  
 \mathchardef\PDelta="7101
 \mathchardef\PXi="7104
 \mathchardef\PLambda="7103
 \mathchardef\PSigma="7106
 \mathchardef\POmega="710A
 \mathchardef\PUpsilon="7107
                  
 \def\PB      {\ensuremath{B}\xspace}                 
                  
 \def\PD      {\ensuremath{D}\xspace}

 \def\PJ      {\ensuremath{J}\xspace}                 
 \def\PK      {\ensuremath{K}\xspace}

 \def\Pb      {\ensuremath{b}\xspace}                 
 \def\Pc      {\ensuremath{c}\xspace}

 \def\Pi      {\ensuremath{i}\xspace}

}

\def\cquark    {\ensuremath{\Pc}\xspace}

\def\bquark    {\ensuremath{\Pb}\xspace}

\def\kaon  {\ensuremath{\PK}\xspace}
  \def\Kbar  {\kern 0.2em\overline{\kern -0.2em \PK}{}\xspace}

\def\Kz    {\ensuremath{\kaon^0}\xspace}
\def\Kzb   {\ensuremath{\Kbar^0}\xspace}
\def\KzKzb {\ensuremath{\Kz \kern -0.16em \Kzb}\xspace}
\def\Kp    {\ensuremath{\kaon^+}\xspace}
\def\Km    {\ensuremath{\kaon^-}\xspace}

\def\KpKm  {\ensuremath{\Kp \kern -0.16em \Km}\xspace}

  \def\Dbar    {\kern 0.2em\overline{\kern -0.2em \PD}{}\xspace}
\def\D       {\ensuremath{\PD}\xspace}

\def\Dz      {\ensuremath{\D^0}\xspace}
\def\Dzb     {\ensuremath{\Dbar^0}\xspace}
\def\DzDzb   {\ensuremath{\Dz {\kern -0.16em \Dzb}}\xspace}
\def\Dp      {\ensuremath{\D^+}\xspace}
\def\Dm      {\ensuremath{\D^-}\xspace}

\def\DpDm    {\ensuremath{\Dp {\kern -0.16em \Dm}}\xspace}

  \def\Bbar    {\kern 0.18em\overline{\kern -0.18em \PB}{}\xspace}

\def\jpsi     {\ensuremath{{\PJ\mskip -3mu/\mskip -2mu\Ppsi\mskip 2mu}}\xspace}

\def\etac     {\ensuremath{\Peta_\cquark}\xspace}

  \def\Y#1S{\ensuremath{\PUpsilon{(#1S)}}\xspace}

\def\Lbar {\ensuremath{\kern 0.1em\overline{\kern -0.1em\PLambda}}\xspace}
\def\Lambdares {\ensuremath{\PLambda}\xspace}

\def\to                 {\ensuremath{\rightarrow}\xspace}

\def\AT#1     {\ensuremath{A_{\mathrm{T}}^{#1}}\xspace}           

\def\C#1      {\ensuremath{\mathcal{C}_{#1}}\xspace}                       
\def\Cp#1     {\ensuremath{\mathcal{C}_{#1}^{'}}\xspace}                    
\def\Ceff#1   {\ensuremath{\mathcal{C}_{#1}^{\mathrm{(eff)}}}\xspace}        
\def\Cpeff#1  {\ensuremath{\mathcal{C}_{#1}^{'\mathrm{(eff)}}}\xspace}       
\def\Ope#1    {\ensuremath{\mathcal{O}_{#1}}\xspace}                       
\def\Opep#1   {\ensuremath{\mathcal{O}_{#1}^{'}}\xspace}                    

\newcommand{\tev}{\ensuremath{\mathrm{\,Te\kern -0.1em V}}\xspace}
\newcommand{\gev}{\ensuremath{\mathrm{\,Ge\kern -0.1em V}}\xspace}
\newcommand{\mev}{\ensuremath{\mathrm{\,Me\kern -0.1em V}}\xspace}
\newcommand{\kev}{\ensuremath{\mathrm{\,ke\kern -0.1em V}}\xspace}
\newcommand{\ev}{\ensuremath{\mathrm{\,e\kern -0.1em V}}\xspace}
\newcommand{\gevc}{\ensuremath{{\mathrm{\,Ge\kern -0.1em V\!/}c}}\xspace}
\newcommand{\mevc}{\ensuremath{{\mathrm{\,Me\kern -0.1em V\!/}c}}\xspace}
\newcommand{\gevcc}{\ensuremath{{\mathrm{\,Ge\kern -0.1em V\!/}c^2}}\xspace}
\newcommand{\gevgevcccc}{\ensuremath{{\mathrm{\,Ge\kern -0.1em V^2\!/}c^4}}\xspace}
\newcommand{\mevcc}{\ensuremath{{\mathrm{\,Me\kern -0.1em V\!/}c^2}}\xspace}

\def\mum  {\ensuremath{\,\upmu\rm m}\xspace}

\newcommand{\chisq}{\ensuremath{\chi^2}\xspace}

\def\gsim{{~\raise.15em\hbox{$>$}\kern-.85em
          \lower.35em\hbox{$\sim$}~}\xspace}
\def\lsim{{~\raise.15em\hbox{$<$}\kern-.85em
          \lower.35em\hbox{$\sim$}~}\xspace}

\def\pt         {\mbox{$p_{\rm T}$}\xspace}

\def\evtgen     {\mbox{\textsc{EvtGen}}\xspace}
\def\pythia     {\mbox{\textsc{Pythia}}\xspace}

\def\geant      {\mbox{\textsc{Geant4}}\xspace}

\def\photos     {\mbox{\textsc{Photos}}\xspace}

\def\tell1  {TELL1\xspace}
\def\ukl1   {UKL1\xspace}

\usepackage{cite} 
\usepackage{mciteplus}

\begin{document}

\renewcommand{\thefootnote}{\fnsymbol{footnote}}
\setcounter{footnote}{1}

\begin{titlepage}
\pagenumbering{roman}

\vspace*{-1.5cm}
\centerline{\large EUROPEAN ORGANIZATION FOR NUCLEAR RESEARCH (CERN)}
\vspace*{1.5cm}
\hspace*{-0.5cm}
\begin{tabular*}{\linewidth}{lc@{\extracolsep{\fill}}r}
\ifthenelse{\boolean{pdflatex}}
{\vspace*{-2.7cm}\mbox{\!\!\!\includegraphics[width=.14\textwidth]{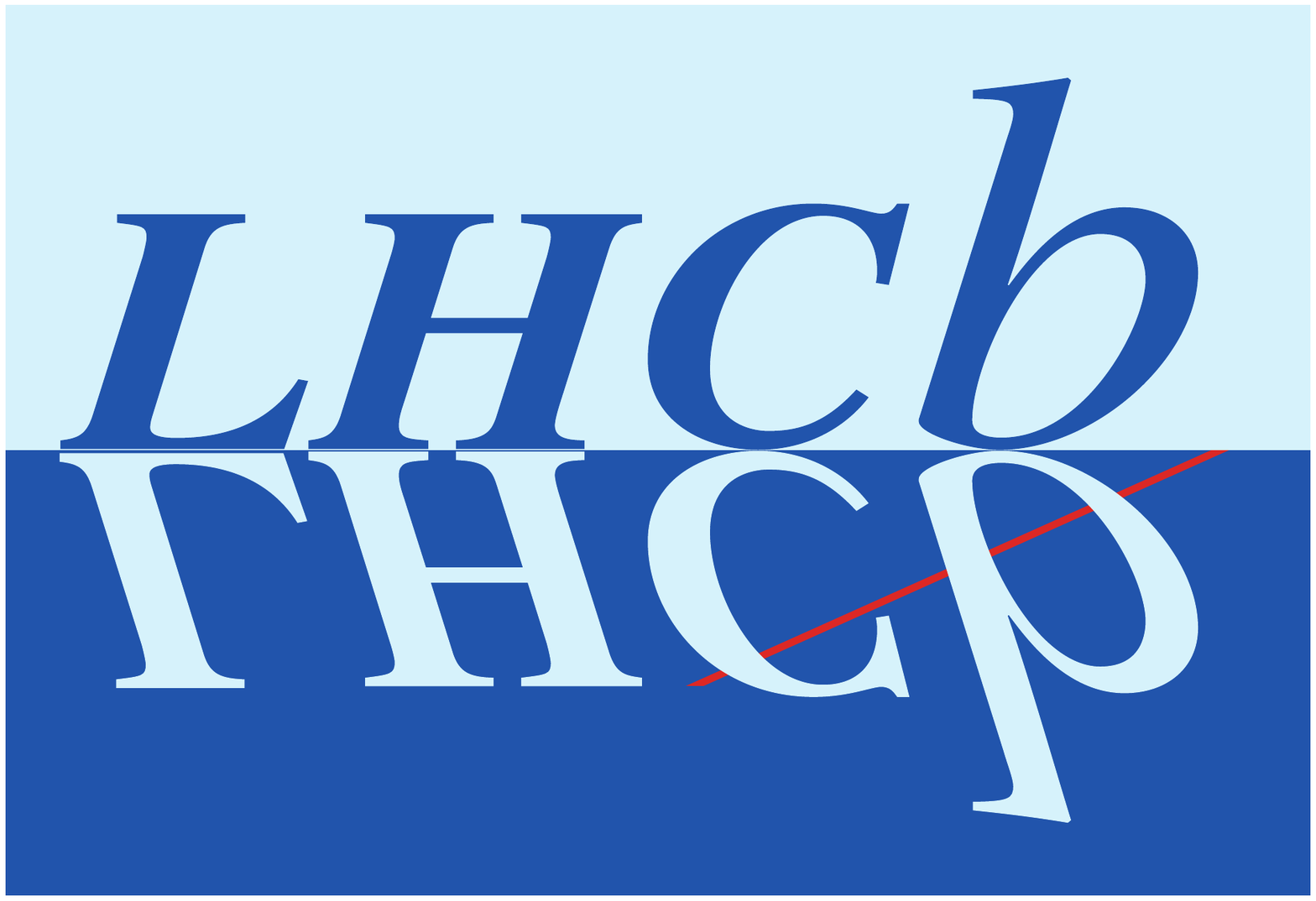}} & &}%
{\vspace*{-1.2cm}\mbox{\!\!\!\includegraphics[width=.12\textwidth]{lhcb-logo.eps}} & &}%
\\
 & & CERN-PH-EP-2013-040 \\  
 & & LHCb-PAPER-2012-047 \\  
 & & 27 March 2013 \\ 
 & & \\
\end{tabular*}

\vspace*{1.0cm}

{\bf\boldmath\huge
\begin{center}
  Measurements of the branching fractions of $B^{+} \to p \bar p
  K^{+}$ decays
\end{center}
}

\vspace*{0.5cm}

\begin{center}
The LHCb collaboration\footnote{Authors are listed on the following pages.}
\end{center}

\vspace{\fill}

\begin{abstract}
  \noindent
The
branching fractions of the decay $\bppk$ for different
intermediate states are measured using data, corresponding to an
integrated luminosity of $1.0\unitm{fb^{-1}}$, collected by the LHCb
experiment.
The total
branching fraction, its charmless component \mbox{$(M_{p\bar p}
<2.85\unitm{GeV/}c^{2})$} and the branching fractions via the resonant $c\bar c$
states $\eta_{c}(1S)$ and $\psi(2S)$ relative to the decay via a \jpsi
intermediate state are
\begin{align*}
\frac{{\mathcal B}(\bppk)_{\rm total}}{{\mathcal B}(B^{+} \to J/\psi K^{+} \to p
\bar p K^{+})}=& \, 4.91 \pm 0.19 \, {(\rm
  stat)} \pm 0.14 \, {(\rm syst)},\\
\frac{{\mathcal B}(\bppk)_{\mpp \unitm{GeV/}c^{2}}}{{\mathcal B}(B^{+} \to J/\psi K^{+} \to p
\bar p K^{+})}=& \, 2.02 \pm 0.10 \, {(\rm
  stat)}\pm 0.08 \, {(\rm syst)},\\
\frac{{\mathcal B} (B^{+} \to \eta_{c}(1S) K^{+} \to p
\bar p K^{+})}{{\mathcal B}(B^{+} \to J/\psi K^{+} \to p
\bar p K^{+})} = & \, 0.578 \pm 0.035 \, {(\rm stat)} \pm 0.027 \, {(\rm syst)},\\
\frac{{\mathcal B} (B^{+} \to \psi(2S)
K^{+} \to p
\bar p K^{+})}{{\mathcal B}(B^{+} \to J/\psi K^{+} \to p
\bar p K^{+})}=& \, 0.080 \pm 0.012 \, {(\rm stat)} \pm  0.009 \, {(\rm syst)}.\\
\end{align*}
Upper limits on the $B^{+}$ branching fractions into the
$\eta_{c}(2S)$ meson and into the charmonium-like states $X(3872)$ and $X(3915)$
are also obtained.
\end{abstract}

\vspace*{0.5cm}

\begin{center}
  Submitted to EPJ C
\end{center}

\vspace{\fill}

{\footnotesize 
\centerline{\copyright~CERN on behalf of the \lhcb collaboration, license \href{http://creativecommons.org/licenses/by/3.0/}{CC-BY-3.0}.}}
\vspace*{2mm}

\end{titlepage}


\newpage
\setcounter{page}{2}
\mbox{~}
\newpage

\centerline{\large\bf LHCb collaboration}
\begin{flushleft}
\small
R.~Aaij$^{38}$, 
C.~Abellan~Beteta$^{33,n}$, 
A.~Adametz$^{11}$, 
B.~Adeva$^{34}$, 
M.~Adinolfi$^{43}$, 
C.~Adrover$^{6}$, 
A.~Affolder$^{49}$, 
Z.~Ajaltouni$^{5}$, 
J.~Albrecht$^{9}$, 
F.~Alessio$^{35}$
% ...

{\footnotesize \it
% ...
$ ^{5}$Clermont Universit\'{e}, Universit\'{e} Blaise Pascal, CNRS/IN2P3, LPC, Clermont-Ferrand, France\\
$ ^{6}$CPPM, Aix-Marseille Universit\'{e}, CNRS/IN2P3, Marseille, France\\
% ...
$ ^{9}$Fakult\"{a}t Physik, Technische Universit\"{a}t Dortmund, Dortmund, Germany\\
% ...
$ ^{11}$Physikalisches Institut, Ruprecht-Karls-Universit\"{a}t Heidelberg, Heidelberg, Germany\\
% ...
$ ^{33}$Universitat de Barcelona, Barcelona, Spain\\
$ ^{34}$Universidad de Santiago de Compostela, Santiago de Compostela, Spain\\
$ ^{35}$European Organization for Nuclear Research (CERN), Geneva, Switzerland\\
% ...
$ ^{38}$Nikhef National Institute for Subatomic Physics, Amsterdam, The Netherlands\\
% ...
$ ^{43}$H.H. Wills Physics Laboratory, University of Bristol, Bristol, United Kingdom\\
% ...
$ ^{49}$Oliver Lodge Laboratory, University of Liverpool, Liverpool, United Kingdom\\
% ...
$ ^{n}$LIFAELS, La Salle, Universitat Ramon Llull, Barcelona, Spain\\
% ...
}
\end{flushleft}

\cleardoublepage

\renewcommand{\thefootnote}{\arabic{footnote}}
\setcounter{footnote}{0}

\pagestyle{plain} 
\setcounter{page}{1}
\pagenumbering{arabic}


\section{Introduction}
\label{sec:Introduction}
The $B^{+} \to p \bar p K^{+}$ decay\footnote{The inclusion of charge-conjugate modes is implied throughout the
paper.} offers a clean environment to study $c\bar c$ states and
charmonium-like mesons that decay to
$p\bar p$ and excited ${\bar \Lambdares}$ baryons that decay to $\bar
p K^{+}$,
and to search
for glueballs or exotic states. The presence of $p \bar p$ in the
final state allows intermediate
states of any quantum numbers to be studied and the existence of the
charged kaon in the final state significantly enhances the signal to
background ratio in the selection procedure.
Measurements of intermediate charmonium-like states, such as the $X(3872)$, are important to clarify their
nature~\cite{Brambilla:2010cs, Aaij:2013zoa} and to determine their partial width to
$p\bar p$, which is crucial
to predict the production rate of these states in dedicated experiments~\cite{Lange:2010dt}. 
BaBar and Belle have previously measured the
$\bppk$ branching fraction, including contributions from the $J/\psi$
and $\eta_{c}(1S)$ intermediate states~\cite{Aubert:2005gw, Wei:2007fg}. 
The data sample, corresponding to an
integrated luminosity of $1.0\unitm{fb^{-1}}$, collected by LHCb at $\sqrt{s}=7\unitm{TeV}$ allows the study of
substructures in the $B^{+}\to p\bar p K^{+}$ decays with a sample ten times
larger than those available at previous experiments.

In this paper we report measurements of the ratios of branching fractions

\begin{equation}
{\cal R}({\rm mode}) =\frac{{\cal B}(B^{+} \to {\rm mode}\to p\bar p
  K^{+})}{{\cal B}(B^{+}\to J/\psi K^{+}\to p\bar p K^{+})},
  \end{equation}
where ``mode'' corresponds to the intermediate $\eta_{c}(1S)$, $\psi(2S)$, $\eta_{c}(2S)$,
$\chi_{c0}(1P)$, $h_{c}(1P)$, $X(3872)$ or $X(3915)$ states, together with
a kaon.

\section{Detector and software}
\label{sec:Detector}
The \lhcb detector~\cite{Alves:2008zz} is a single-arm forward
spectrometer covering the \mbox{pseudorapidity} range $2<\eta <5$,
designed for the study of particles containing \bquark or \cquark quarks. The
detector includes a high precision tracking system consisting of a
silicon-strip vertex detector surrounding the $pp$ interaction region,
a large-area silicon-strip detector located upstream of a dipole
magnet with a bending power of about $4{\rm\,Tm}$, and three stations
of silicon-strip detectors and straw drift-tubes placed
downstream. The combined tracking system has momentum $(p)$ resolution
$\Delta p/p$ that varies from 0.4\% at 5\gevc to 0.6\% at 100\gevc,
and impact parameter resolution of 20\mum for tracks with high
transverse momentum (\pt). Charged hadrons are identified using two
ring-imaging Cherenkov (RICH) detectors. Photon, electron and hadron
candidates are identified by a calorimeter system consisting of
scintillating-pad and pre-shower detectors, an electromagnetic
calorimeter and a hadronic calorimeter. Muons are identified by a
system composed of alternating layers of iron and multiwire
proportional chambers.

The trigger~\cite{Aaij:2012me} consists of a
hardware stage, based on information from the calorimeter and muon
systems, followed by a software stage where candidates are fully
reconstructed.
The hardware trigger selects hadrons with high transverse energy
in the calorimeter.
The software trigger requires a two-, three- or four-track
secondary vertex with a high \pt sum of
the tracks and a significant displacement from the primary $pp$
interaction vertices~(PVs). At least one track should have $\pt >
1.7\gevc$ and impact parameter~(IP) \chisq with respect to the
primary interaction greater than 16. The IP \chisq is defined as the
difference between the \chisq of the PV reconstructed with and
without the considered track. A multivariate algorithm is
used for the identification of secondary vertices consistent with the
decay of a \bquark hadron.

Simulated $B^{+} \to p \bar p K^{+}$ decays, generated uniformly in phase space, are used to optimize the signal selection and to evaluate
the ratio of the efficiencies for each considered channel with
respect to the $J/\psi$ channel. Separate samples of $B^{+} \to J/\psi K^{+} \to
p \bar p K^{+}$ and $B^{+} \to\eta_{c}(1S) K^{+} \to
p \bar p K^{+}$ decays, generated with the known angular distributions,
are used to check the dependence of the efficiency ratio on the
angular distribution.
In the simulation, $pp$ collisions are generated using
\pythia~6.4~\cite{Sjostrand:2006za} with a specific \lhcb
configuration~\cite{LHCb-PROC-2010-056}.  Decays of hadronic particles
are described by \evtgen~\cite{Lange:2001uf} in which final state
radiation is generated by \photos~\cite{Golonka:2005pn}. The
interaction of the generated particles with the detector and its
response are implemented using the \geant
toolkit~\cite{Allison:2006ve, *Agostinelli:2002hh} as described in
Ref.~\cite{LHCb-PROC-2011-006}.

\section{Candidate selection}
\label{sec:Selection}
Candidate $B^{+}\to p\bar p K^{+}$ decays are reconstructed from any combination of three charged tracks with total charge of $+1$.
The final state particles are required to have a track fit with a $\chi^{2}/{\rm ndf} < 3$ where ndf is the number of degrees of freedom. They must also have $p > 1500\unitm{MeV/}c$, \pt $> 100\unitm{MeV}/c$, and IP $\chi^2>1$ with respect to any primary vertex in the event. 
Particle identification (PID) requirements, based
on the RICH detector information, are applied to $p$ and $\bar p$ candidates. The discriminating variables between different particle hypotheses ($\pi$, $K$, $p$) are the differences between log-likelihood values $\Delta$ln${\mathcal L}_{\alpha \beta}$ under particle hypotheses $\alpha$ and $\beta$, respectively.
The $p$ and $\bar p$ candidates are required to have $\Delta$ln${\mathcal L}_{p\pi}>-5$.
The reconstructed $B^{+}$ candidates are required to have an invariant mass in the range $5079-5579\unitm{MeV/}c^{2}$. The asymmetric invariant mass range around the nominal $B^{+}$ mass is designed to select also $B^{+} \to p \bar p \pi^{+}$ candidates without any requirement on the PID of the kaon. The PV associated to each $B^{+}$ candidate is defined to be the one for which the $B^{+}$ candidate has the smallest IP $\chi^{2}$. 
The $B^{+}$ candidate is required to
have a vertex fit with a $\chi^{2}/{\rm ndf}<12$ and a distance greater than $3\unitm{mm}$, a $\chi^{2}$ for the flight distance greater than $500$, and an IP $\chi^2<10$ with respect to the associated PV. The maximum distance of closest approach between daughter tracks has to be less than $0.2\unitm{mm}$. The angle between the reconstructed momentum of the $B^{+}$ candidate and the $B^{+}$ flight direction ($\theta_{\rm fl}$) is required to have $\cos\theta_{\rm fl}>0.99998$. 

\begin{figure}
\centering
\includegraphics[scale=0.5] {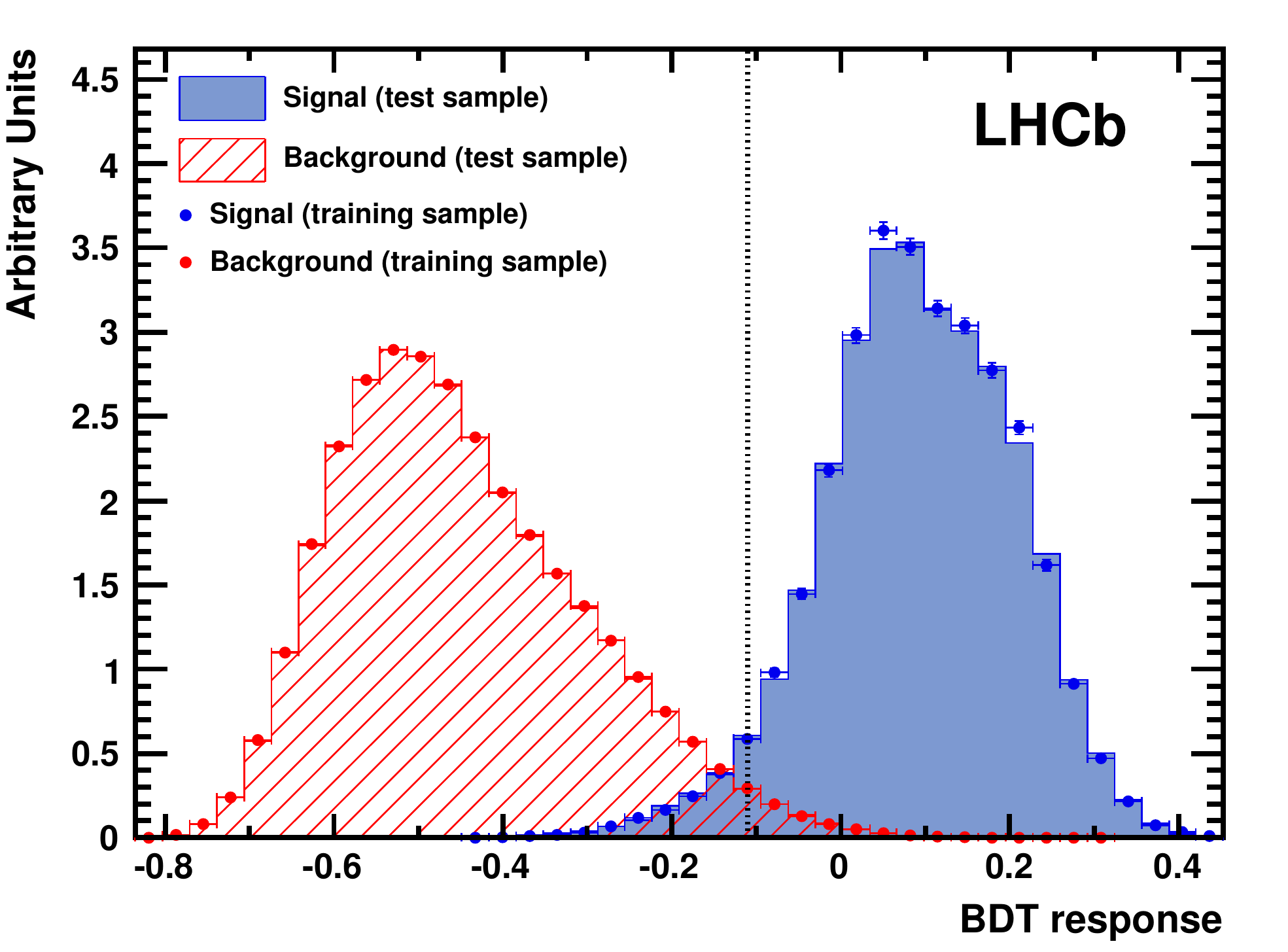}
\caption{\small Distribution of the BDT algorithm response evaluated for
  background candidates from the data sidebands (red), and signal candidates from simulation (blue). The black dotted line indicates the chosen BDT response value.}
\label{fig:bdtoutput}
\end{figure}

The reconstructed candidates that meet the above criteria are filtered using a boosted decision tree
(BDT) algorithm~\cite{Breiman}. The BDT is trained with a sample of simulated $B^{+} \to p
\bar p K^{+}$ signal candidates and a background sample of data candidates taken from the invariant mass sidebands in the ranges
$5080-5220 \unitm{MeV/}c^{2}$ and $5340-5480 \unitm{MeV/}c^{2}$.
The variables used by the BDT to discriminate between signal and background candidates are: the \pt of each reconstructed track; the sum of the daughters' $p_{T}$; the
sum of the \hbox{IP $\chi^{2}$} of the three daughter tracks with respect to the primary vertex; the
IP of the daughter, with the highest \pt, with respect to the primary vertex; the number of daughters with \pt$> 900\unitm{GeV/}c$; the maximum distance
of closest approach between any two of the $B^{+}$ daughter particles; the IP of the $B^{+}$ candidate with respect to the primary vertex; the distance between primary and secondary vertices; the $\theta_{\rm fl}$ angle; the $\chi^{2}/{\rm ndf}$ of the secondary vertex; a pointing variable defined as $\frac{P\sin\theta}{P\sin\theta + \sum_{i}
    p_{\rm T,i}}$, where $P$ is the total momentum of the three-particle final state, $\theta$ is the angle between the direction of the sum of
the daughter's momentum and the direction of the flight
distance of the $B^{+}$ and $\sum_{i} p_{{\rm T},i}$ is the sum of the
transverse momenta of the daughters; and the log likelihood difference for each daughter between the assumed PID hypothesis and the pion hypothesis.
The selection criterion on the BDT response (Fig.~\ref{fig:bdtoutput}) is chosen in order to have a signal to background ratio of the order of unity. This corresponds to a BDT response value of $-0.11$. The efficiency of the BDT selection is greater than $92\%$ with a background rejection greater than $86\%$.

\begin{figure}[htb!!]
\centering
\includegraphics [scale=0.45]{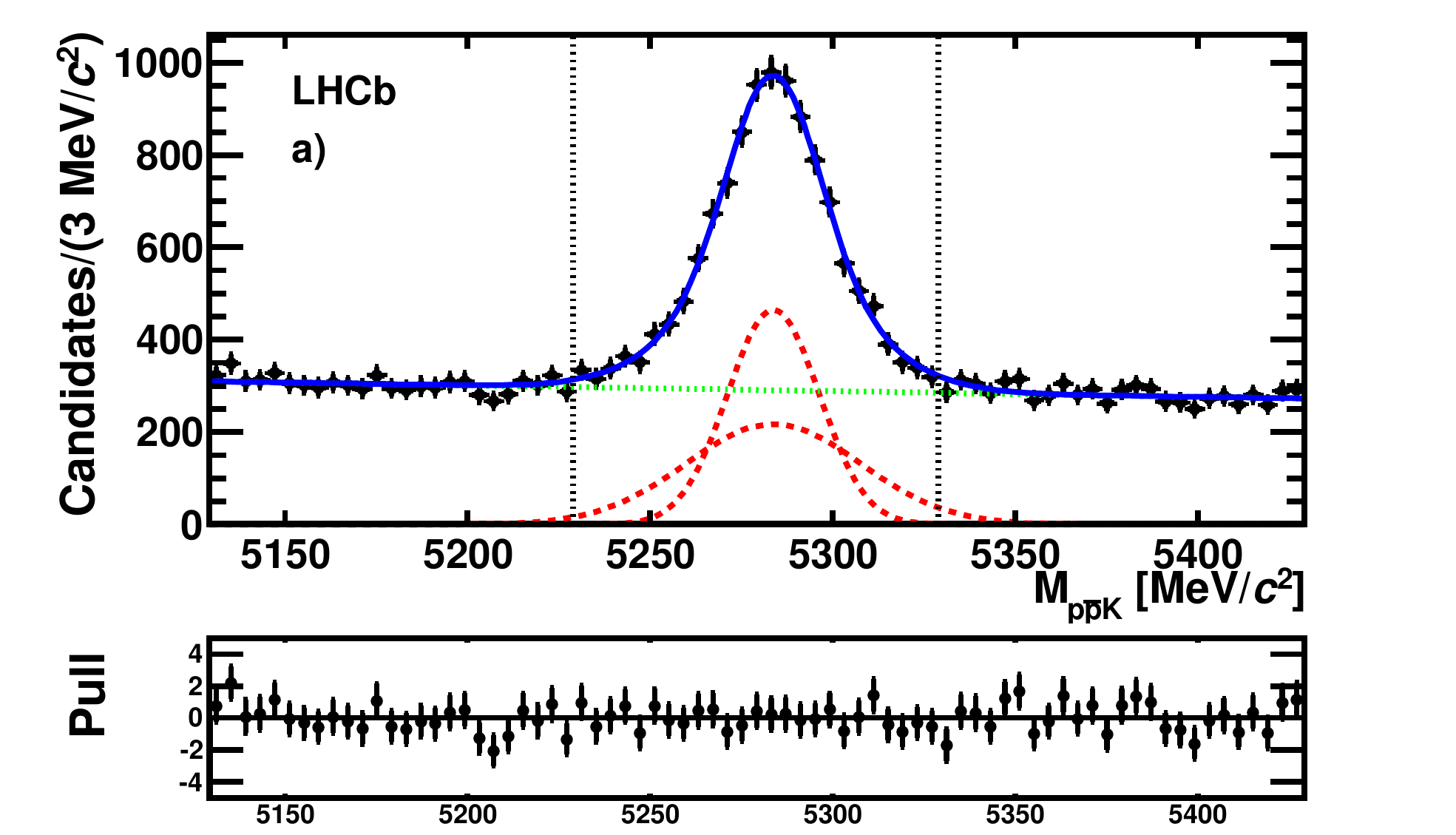}\\[3pt]
\includegraphics [scale=0.45]{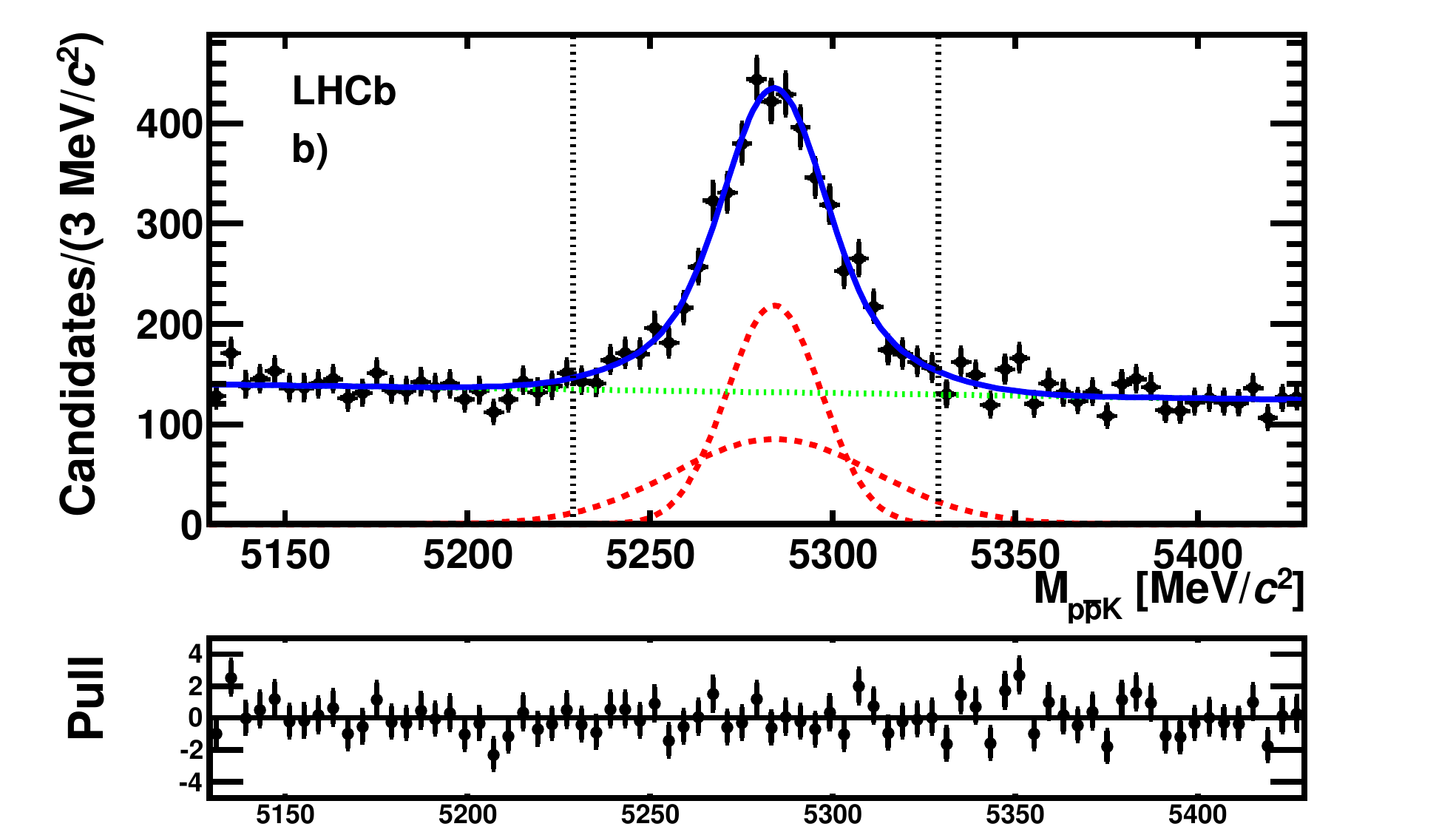}
\caption{\small Invariant mass distribution of a) all selected $\bppk$ candidates and b) candidates having \mbox{$\mpp \unitm{GeV/}c^{2}$}. The points with error bars are the data and the solid lines are the result of the fit. The dotted lines represent the two Gaussian functions (red) and the dashed line the linear function (green) used to parametrize the signal and the background, respectively. The vertical lines indicate the signal region. The two plots below the mass distributions show the pulls.}
\label{fig:all285}
\end{figure} 
\section{Signal yield determination}
\label{sec:massfits}

\begin{figure}[htb]
\centering
\includegraphics [scale=0.8]{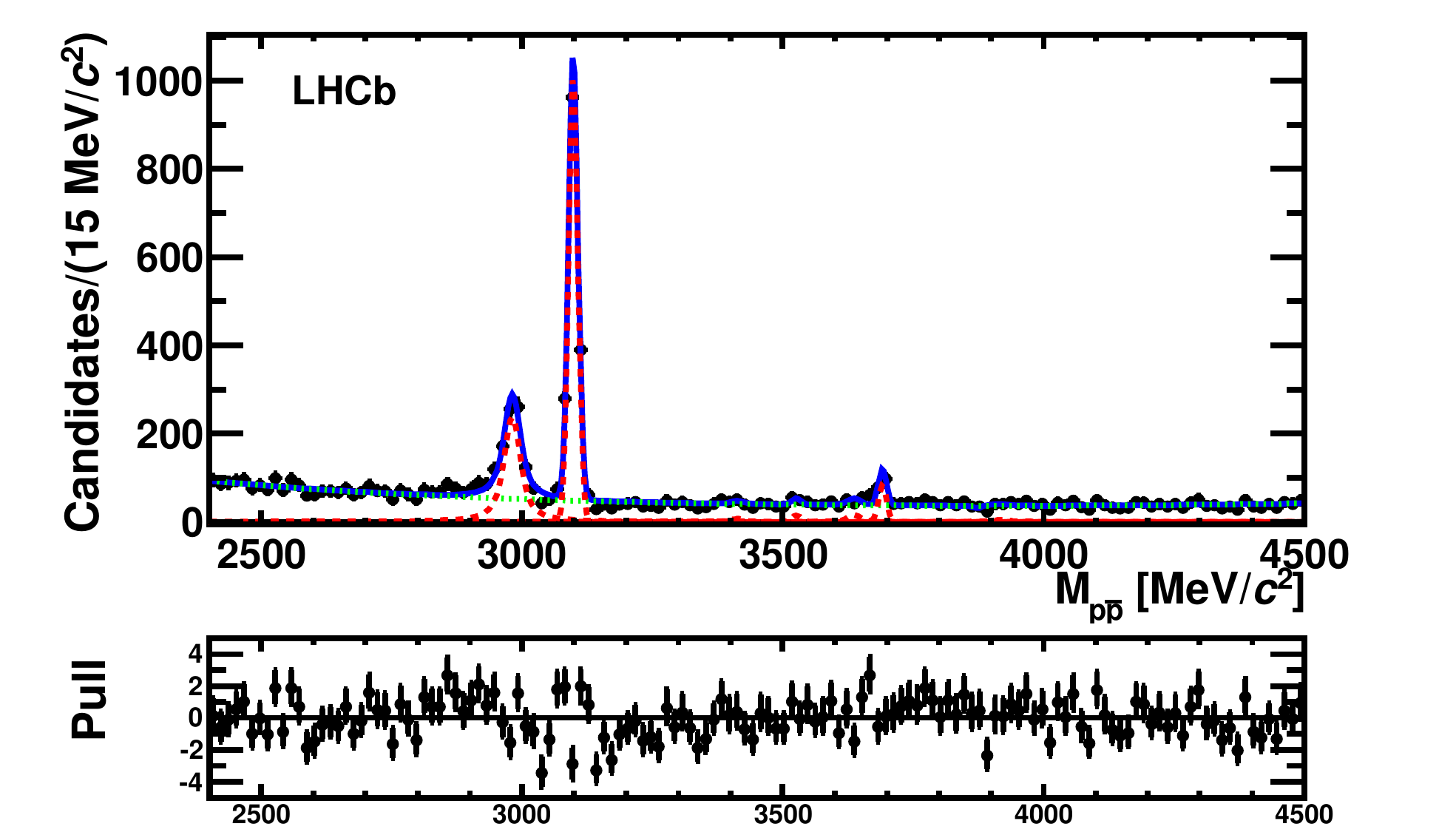}
\caption{
\small Invariant mass distribution of the $p \bar p$ system for 
$B^{+} \to p\bar p K^{+}$ candidates within the $B^{+}$ mass signal window, $\vert M(p\bar p K^{+}) - M_{B^+}\vert < 50 \unitm{MeV/}c^{2}$. The dotted lines represent the Gaussian and Voigtian functions (red) and the dashed line the smooth function (green) used to parametrize the signal and the background, respectively.
The bottom plot shows the pulls.
}
\label{fig:cc}
\end{figure} 

\begin{figure}[htb]
\centering
\includegraphics [scale=0.8]{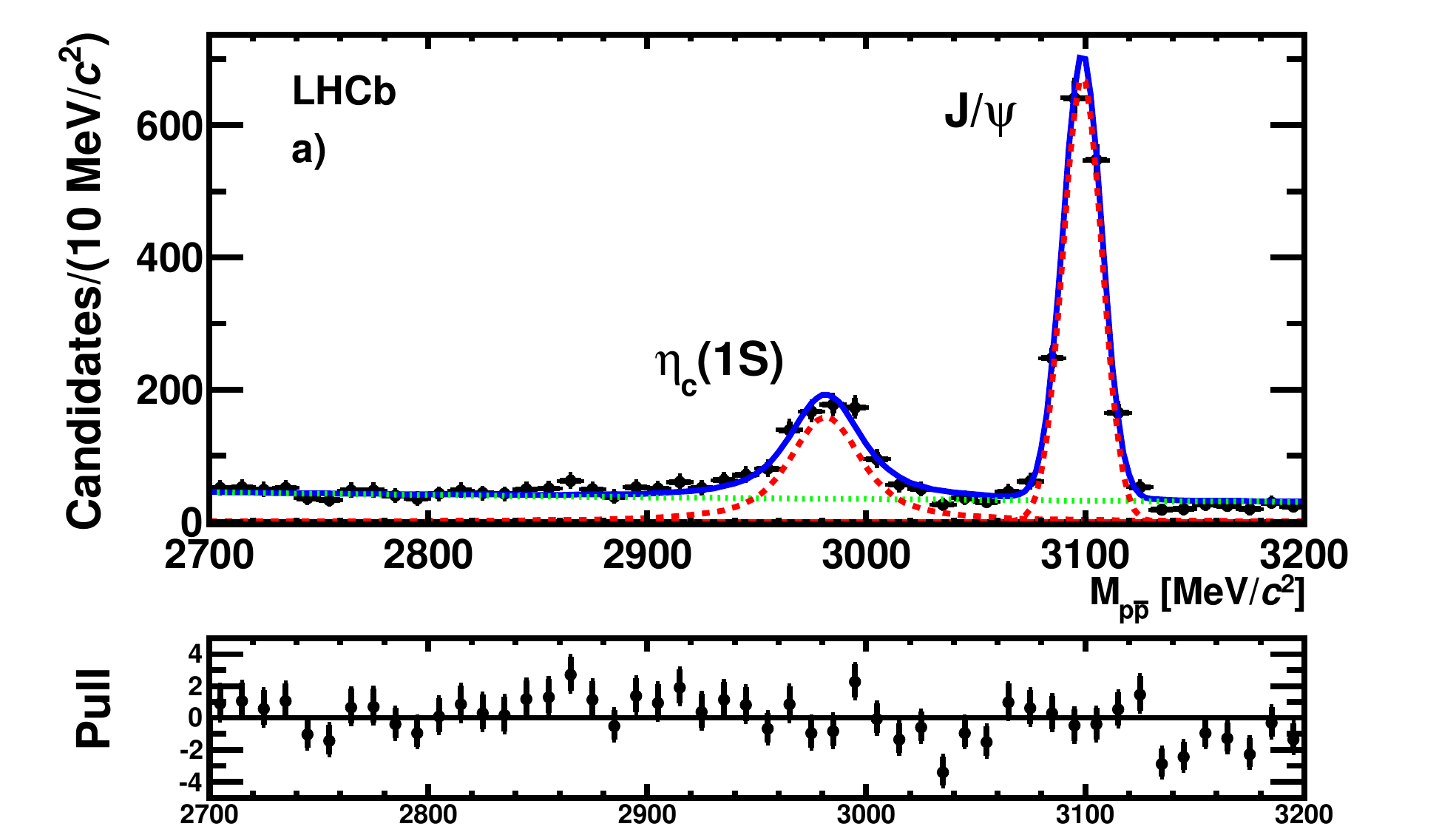}
\includegraphics [scale=0.8]{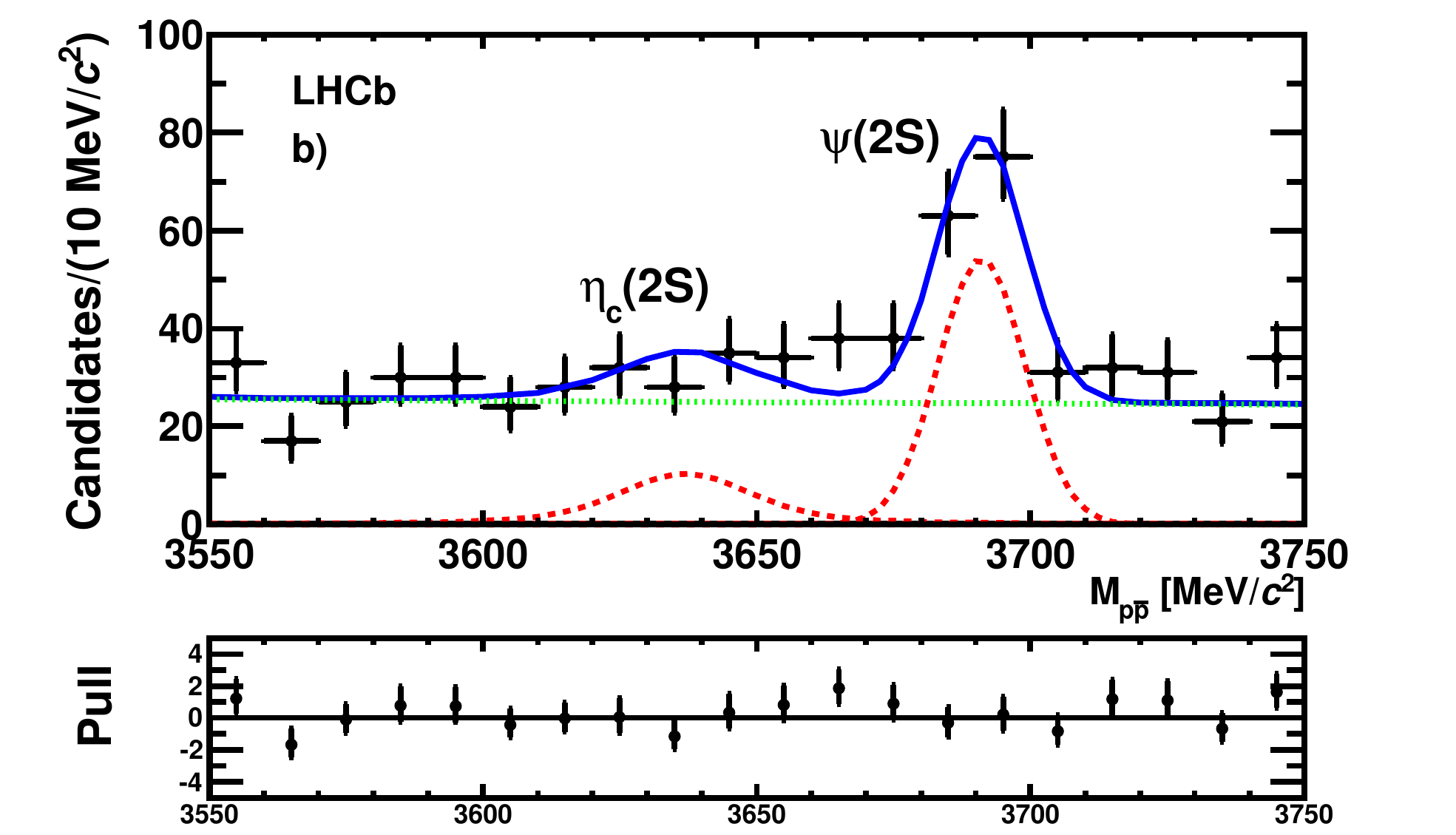}
\caption{
\small Invariant mass distribution of the $p \bar p$ system in the regions around a) the $\eta_{c}(1S)$ and $J/\psi$ and b) the $\eta_{c}(2S)$ and $\psi(2S)$ states. The dotted lines represent the Gaussian and the Voigtian functions (red) and the dashed line the smooth function (green) used to parametrize the signal and the background, respectively. The two plots below the mass distribution show the pulls.}
\label{fig:ccZOOM}
\end{figure} 

\begin{figure}[htb]
\centering
\includegraphics [scale=0.8]{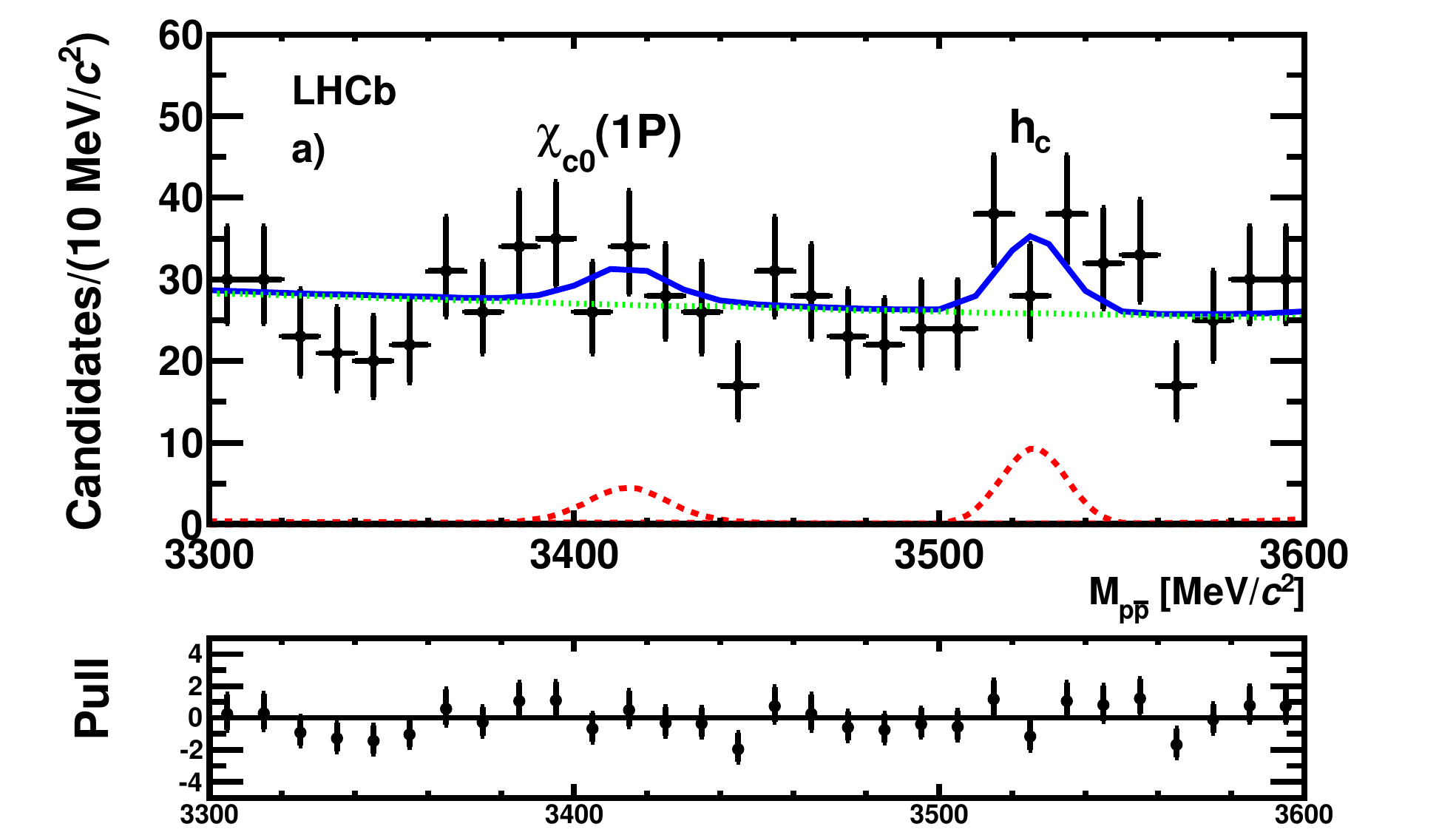}
\includegraphics [scale=0.8]{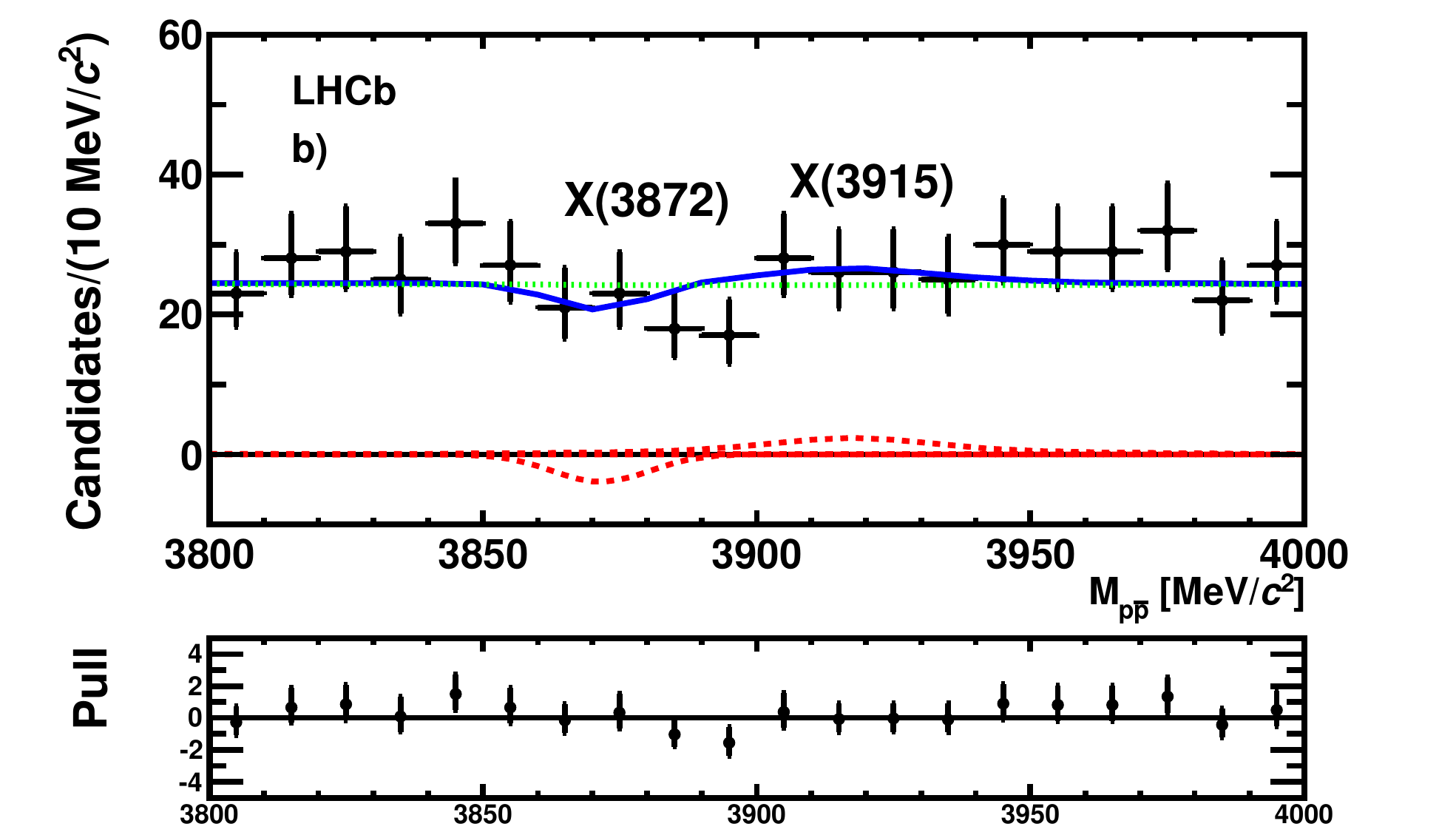}
\caption{\small Invariant mass distribution of the $p \bar p$ system in the regions around a)
the $\chi_{c0}(1P)$ and $h_{c}$ and b) the $X(3872)$ and $X(3915)$ states. The dotted lines represent the Gaussian and Voigitian functions (red) and the dashed line the smooth function (green) used to parametrize the signal and the background, respectively. The two plots below the mass distribution show the pulls.}
\label{fig:ccZOOM1}
\end{figure} 

The signal yield is determined from an unbinned extended maximum likelihood fit to the invariant mass
of  selected $B^{+} \to p \bar p K^{+}$ candidates, shown in Fig.~\ref{fig:all285}a). The signal component is parametrized as the sum of two Gaussian functions with the same mean and different widths. The background component is parametrized as a linear function.
The signal yield of the charmless component is determined 
by performing the same fit described above to the sample of
$B^{+} \to p \bar p K^{+}$ candidates with $M_{p\bar p} < 2.85 \unitm{GeV/}c^{2}$, shown in Fig.~\ref{fig:all285}b). 
The $B^{+}$ mass and widths, evaluated with the invariant mass fits to all of the $\bppk$ candidates,
are compatible with the values obtained for the charmless component.

The signal yields for the charmonium contributions, $B^{+}\to (c\bar c) K^{+} \to p\bar p K^{+}$, are determined by fitting the
$p\bar p$ invariant mass distribution of $B^{+} \to p\bar p K^{+}$ candidates within the $B^{+}$ mass signal window, $\vert M_{p\bar p K^{+}} - M_{B^+}\vert < 50 \unitm{MeV/}c^{2}$. Simulations show that no narrow structures are induced in the $p \bar p$ spectrum as kinematic reflections of possible $B^{+} \to p {\it \bar \Lambda} \to p \bar p K^{+}$ intermediate states. 

An unbinned extended maximum likelihood fit to the 
$p\bar p$ invariant mass distribution, shown in Fig.~\ref{fig:cc}, is performed over the mass range $2400-4500 \unitm{MeV/}c^{2}$. 
The signal components of the narrow resonances $J/\psi$, $\psi(2S)$, $h_c(1P)$, and $X(3872)$, whose natural widths are much smaller than the $p\bar p$ invariant mass resolution, are parametrized by Gaussian functions. 
The signal components for the $\eta_c(1S)$, $\chi_{c0}(1P)$, $\eta_c(2S)$, and $X(3915)$ are parametrized by Voigtian functions.\footnote{A Voigtian function is the convolution of a Breit-Wigner function with a Gaussian distribution.}
Since the $p\bar p$ invariant mass resolution is approximately constant in the explored range, the resolution parameters for all resonances, except the $\psi(2S)$, are fixed to the  $J/\psi$ value ($\sigma_{J/\psi}=8.9 \pm 0.2 \unitm{MeV/}c^{2}$).
The background shape is parametrized as $f(M)=e^{c_{1}M+ c_{2}M^{2}}$ where $c_{1}$ and $c_{2}$ are fit parameters. 
The $J/\psi$ and $\psi(2S)$ resolution parameters, the mass values of the $\eta_c(1S)$, $J/\psi$, and $\psi(2S)$ states, and the $\eta_c(1S)$ natural width are left free in the fit. The masses and widths for the other signal components are fixed to the corresponding world averages~\cite{Beringer:1900zz}. 
The $p\bar p$ invariant mass resolution, determined by the fit to the $\psi(2S)$ is $\sigma_{\psi(2S)}=7.9 \pm 1.7 \unitm{MeV/}c^{2}$.

The fit result is shown in Fig.~\ref{fig:cc}. Figures~\ref{fig:ccZOOM} and~\ref{fig:ccZOOM1} show the details of the fit result in the regions around the $\eta_{c}(1S)$ and $J/\psi$, $\eta_{c}(2S)$ and $\psi(2S)$,
$\chi_{c0}(1P)$ and $h_{c}$, and $X(3872)$ and $X(3915)$ resonances.
Any bias introduced by the inaccurate description of the tails of the  $\eta_{c}(1S)$, $J/\psi$ and $\psi(2S)$ resonances is taken into account in the systematic uncertainty evaluation.

\clearpage

The contribution of $c\bar c\to p\bar p$ from processes other than $\bppk$ decays, denoted as ``non-signal'', is estimated from a fit to the $p \bar p$ mass in the $B^{+}$ mass sidebands $5130-5180$ and $5380 - 5430 \unitm{MeV/}c^{2}$. Except for the $J/\psi$ mode, no evidence of a non-signal contribution is found. The non-signal contribution to the $J/\psi$ signal yield in the $B^{+}$ mass window is $43\pm 11$ candidates and is subtracted from the number of $J/\psi$ signal candidates.

The signal yields, corrected for the non-signal contribution, are reported in Table~\ref{tab:yield}. For the intermediate charmonium states $\eta_{c}(2S)$, $\chi_{c0}(1P)$,
$h_{c}(1P)$, $X(3872)$ and $X(3915)$, there is no evidence of signal. The $95\% \unitm{CL}$ upper
limits on the number of candidates are shown in Table~\ref{tab:yield} and are determined from the
likelihood profile integrating over the nuisance parameters. Since for the $X(3872)$ the fitted signal yield is negative, the upper limit has been calculated integrating the likelihood only in the physical region of a signal yield greater than zero.

\begin{table}[tb]
\caption{\small Signal yields for the different channels and corresponding 95\% 
\unitm{CL} upper limits for modes with less than 3$\sigma$ statistical significance. For the $J/\psi$ mode, the non-signal yield is subtracted. Uncertainties are statistical only.
}
\renewcommand{\arraystretch}{1.1}
\vskip 0.3cm
\centering
\begin{tabular}{l|r@{$\pm$}l|c}
 $B^{+}$ decay mode & \multicolumn{2}{c|}{Signal yield} & Upper limit  (95\%  \unitm{CL})\\
\hline
$p\bar p K^{+}\; {\rm [total]}$ & $6951\,$&$\,176$&\\
$p\bar p K^{+}\;[\mpp \unitm{GeV/}c^{2}] $ & $3238\,$ &$\,122$ &\\
$\jpsi K^{+}$ & $1458\,$ & $\,42$ & \\
$\eta_{c}(1S) K^{+}$ & $856\,$ & $\,46$ & \\
$\psi(2S)K^{+}$ & $107\,$ & $\,16$& \\
$\etac(2S)K^{+}$ & $39\,$ & $\,15$ & $< 65.4$\\
$\chi_{c0}(1P)K^{+}$ & $15\,$ & $\,13$ & $<38.1$ \\
$h_{c}(1P)K^{+}$ & $21\,$ & $\,11$ & $<40.2$\\
$X(3872)K^{+}$ & $-9\,$ & $\,8$ & $<10.3$ \\
$X(3915)K^{+}$ & $13\,$ & $\,17$ & $<42.1$\\
\end{tabular}
\label{tab:yield}
\end{table}

\section{Efficiency determination} 
\label{sec:eff}
The ratio of branching fractions is calculated using
\begin{equation}
{\cal R}({\rm mode}) =\frac{{\cal B}(B^{+} \to {\rm mode}\to p\bar
  p K^{+})}{{\cal B}(B^{+}\to J/\psi K^{+}\to p\bar p K^{+})} = \frac{N_{\rm mode}}{N_{\jpsi}}\times
\frac{\epsilon_{\jpsi}}{\epsilon_{\rm mode}},\label{eq:br}
\end{equation} 
where $N_{\rm mode}$ and $N_{\jpsi}$ are the signal yields for the
given mode and the reference mode, $B^{+}\to\jpsi K^{+}\to p\bar p
K^{+}$, and $\epsilon_{\rm mode}/\epsilon_{\jpsi}$ is the corresponding ratio of efficiencies.
The efficiency is the product of the reconstruction,
trigger, and selection efficiencies, and is estimated using
simulated data samples.

\begin{figure}[htb]
\centering
\includegraphics [scale=0.5] {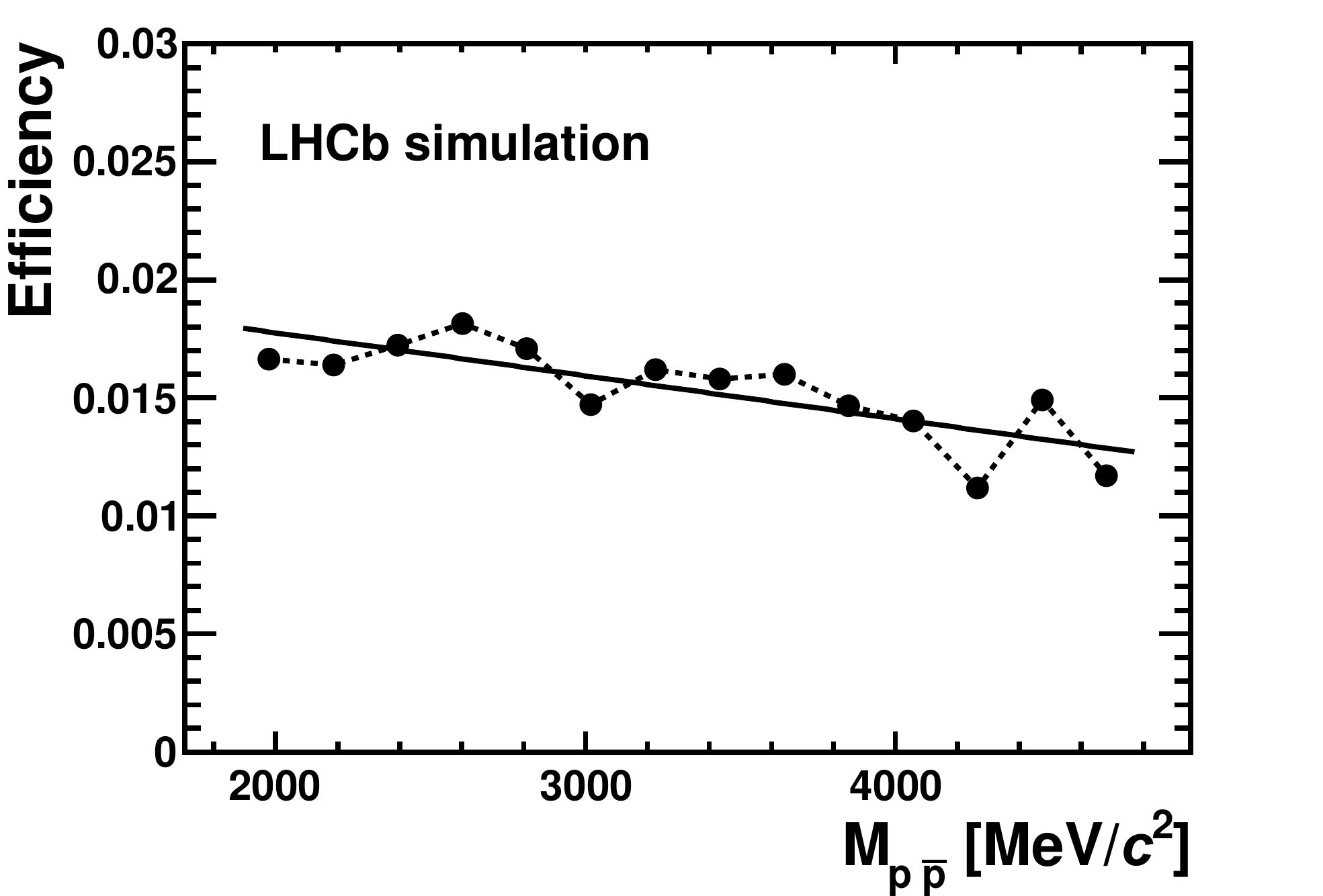}
\caption{\small Efficiency  as a function of $M_{p \bar p}$ for $B^{+}
  \to p \bar p K^{+}$ decays. The solid line represents the linear fit
to the efficiency distribution; the dashed line is the point-by-point
interpolation used to estimate the systematic uncertainty.}
\label{fig:eff}
\end{figure}

Since the track multiplicity distribution for simulated events differs
from that observed in data, simulated candidates are assigned a weight
so that the weighted distribution reproduces the observed
multiplicity distribution.
The distributions of $\Delta$ln${\mathcal L}_{K\pi}$ and
$\Delta$ln${\mathcal L}_{p\pi}$ for kaons and protons in data are
obtained in bins of momentum, pseudorapidity and number of tracks from
control samples of $D^{*+} \to D^{0} (\to K^{-} \pi^{+}) \pi^{+}$
decays for kaons and ${\it \Lambdares} \to p \pi^{-}$ decays for
protons, which are then used on a track-by-track basis to correct the simulation.
The efficiency as a function of $M_{p \bar p}$
is shown in Fig.~\ref{fig:eff}. A linear fit to the efficiency
distribution is performed and the efficiency ratios are determined based on the fit result.

\begin{table}[htbp]
\caption{\small Relative systematic uncertainties (in $\%$) on the
  relative branching fractions from different sources. The total
  systematic uncertainty is determined by adding the individual
  contributions in quadrature.}
\renewcommand{\arraystretch}{1.1}
\vskip 0.3cm
\centering
\small
\begin{tabular}{l|c|c|c|c}
Source & ${\cal R}({\rm total})$ & ${\cal R}(\mpp \unitm{GeV/}c^{2}) $ &
${\cal R}(\etac(1S))$ & ${\cal R}(\psi(2S))$\\
\hline
Efficiency ratio & 0.21 & 0.5 & 3.3 & 4.8\\
\hline
$B^{+}$ mass fit range & 0.16 & 0.5 & $-$ & $-$\\
Sig. and Bkg. shape & 2.5 & 3.6 & 1.8 & 6.5\\
$B^{+}$ mass window & 0.6 & 0.6 & 0.9 & 3.8\\
Non-signal component & $-$  & $-$ & 0.4 & 5.1\\
Signal tail param.& 1.0 & 1.0 & 1.2 & 4.3\\
\hline 
Total & 2.8 & 3.8 & 4.1& 11.3\\
\end{tabular}

\vskip 0.5cm

\begin{tabular}{l|c|c|c|c|c}
Source & ${\cal R}(\eta_{c}(2S))$ &${\cal R}(\chi_{c0}(1P))$ &
${\cal R}(h_{c}(1P))$ & ${\cal R}(X(3872))$ & ${\cal R}(X(3915))$\\
\hline
Efficiency ratio & 4.4 & 2.5 & 3.4 & 6.5 & 7.0\\
\hline
$B^{+}$ mass fit range & $-$ & $-$ & $-$ & $-$& \\
Sig. and Bkg. shape & 3.9 & 3.3 & 14.3 & 5.6 & 10.1 \\
$B^{+}$ mass window & 11.3 & 23.6 & 23.6 & 17.5 & 7.5 \\
Non-signal component & $-$  & $-$ & $-$ & $-$ &  $-$\\
Signal tail param.& 1.0 & 1.0 & 1.0 & 1.0 & 1.0\\
\hline 
Total & 12.8 & 24.0 & 27.8& 19.5 & 15.5\\
\end{tabular}
\label{tab:summarysyst}
\end{table}

\section{Systematic uncertainties}
\label{sec:syst}
The measurements of the relative branching fractions depend on the
ratios of signal yields and efficiencies with respect to the reference
mode. Since the final state is the same in all cases, most of the
systematic uncertainties cancel. 
The systematic uncertainty on the efficiency ratio, in each region of $p
\bar p$ invariant mass, is determined from the difference between the
efficiency ratios calculated using the
solid fitted line and the dashed point-by-point interpolation shown in
Fig.~\ref{fig:eff}.
The uncertainty associated with the evaluation of the $B^{+}$ signal yield has been
determined by varying the fit range by $\pm 30\unitm{MeV/}c^{2}$,
using a single Gaussian instead of a double Gaussian function to model the signal PDF, and using an exponential function to model the background.
For each charmonium
resonance the systematic uncertainty on the signal yield has been investigated
by varying the $B$ mass signal window by $\pm 10\unitm{MeV/}c^{2}$,
the signal and background shape parametrization and the subtraction of
the $c\bar c$ contribution from the continuum.
The systematic uncertainty associated with the parametrization of the signal tails of the $J/\psi$, $\eta_{c}(1S)$ and
$\psi(2S)$ resonances is taken into account by
taking the difference between the number of candidates in the observed
distribution and the number of candidates calculated from the integral of the fit function
in the range $-6\sigma$ to  $-2.5\sigma$.
The systematic uncertainty associated with the selection procedure is
estimated by changing the value of the BDT selection to $-0.03$, which
retains $85\%$ of the signal with a $30\%$ background, and is found to be negligible.
The contributions to the systematic uncertainties from the different sources are 
listed in Table~\ref{tab:summarysyst}. The total systematic
uncertainty is determined by adding the individual contributions in quadrature.

\section{Results}
\label{sec:Results}
The results are summarized in Table~\ref{tab:end} and the values of the product of branching fractions derived from our measurement using the
world average values ${\cal B}(B^{+} \to J/\psi K^{+})
=(1.013\pm0.034)\times 10^{-3}$ and ${\cal B}(J/\psi \to p\bar p)
=(2.17\pm0.07)\times 10^{-3}$~\cite{Beringer:1900zz} are listed in Table~\ref{tab:2011}.

\begin{table}[htb!!!]
\caption{\small Signal yields, efficiency ratios, ratios of branching
  fractions and corresponding upper limits.}
\renewcommand{\arraystretch}{1.1}
\vskip 0.3cm
\centering
\small\begin{tabular}{l|r@{$\pm$}r@{$\pm$}r|c|r@{$\pm$}r@{$\pm$}r|c}
\hline
$B^{+}\to ({\rm mode})$ &  \multicolumn{3}{c|}{Yield} & $\epsilon_{\rm mode}/\epsilon_\jpsi$ & \multicolumn{3}{c|}{${\cal
  R}({\rm mode})$} & Upper Limit\\
~~~~~$\to p\bar p K^{+}$ &  \multicolumn{3}{c|}{$\pm$ stat $\pm$ syst}  & $\pm$ syst  & \multicolumn{3}{c|}{$\pm$ stat $\pm$ syst} &  $95\%$ CL\\
\hline
$\jpsi K^{+}$& $1458\,$ & $\,42\,$ &  $\,24$& $-$ & \multicolumn{3}{c|}{1} & $-$\\
total  & $6951\,$ & $\,176\,$ & $\,171$& $0.970 \pm 0.002$ &
$4.91\, $& $\, 0.19\,$& $\, 0.14 $ & $-$\\
{\scriptsize ${\mpp \unitm{GeV/}c^{2}}$}  & $3238\,$ &$\,122\,$ &$\,121$& $1.097 \pm0.006$&
$2.02\,$& $\,  0.10\,$& $\, 
0.08 $& $-$ \\ 
$\eta_{c}(1S) K^{+}$& $856\,$ &$\,46\,$ & $\,19$& $1.016 \pm
0.034$ &
$0.578\, $& $\,  0.035\, $& $ 
\,0.026$ & $-$\\ 
$\psi(2S)K^{+}$ & $107\,$ & $\,16 $ &$\,13$& $0.921 \pm 0.044$ & $0.080\,$& $\,  0.012\,
$& $ \, 0.009$ & $-$\\
$\etac(2S)K^{+}$ & $39\,$ &$\,15\,$ & $\,5$& $0.927 \pm
0.041$ &
$0.029 \,$& $\,  0.011 \,$& $\,  0.004$ &  $<0.048$\\ 
$\chi_{c0}(1P)K^{+}$ & $15\, $ &$\, 13\, $ &$\,  4$& $0.957 \pm 0.024$ & $0.011 \,$& $\, 
0.009 \,$& $\,  0.003$ & $<0.028$ \\
$h_{c}(1P)K^{+}$& $21\,$ & $\,11\, $ & $\, 5$& $0.943 \pm 0.032$ & $0.015 \,$& $ \,
0.008 \,$& $ \, 0.004$ & 
$<0.029$\\
$X(3872)K^{+}$ & $-9\,$ & $ \,8\, $ & $\,  2$& $0.896 \pm 0.058$ & $-0.007 \,$& $\, 
0.006 \,$& $ \, 0.002$ & $<0.008$\\
$X(3915)K^{+}$ & $13\, $ & $\, 17 \,$ & $ \,5$& $0.890 \pm 0.062$ & $0.010 \,$& $\, 
0.013\, $& $\,  0.002$ & $<0.032$\\
\end{tabular}
\label{tab:end}
\end{table}

\begin{table}[htb!!!]
\caption{\small Branching fractions for $B^{+}\to ({\rm mode})\to p\bar p
  K^{+}$ derived using the
world average value of the ${\cal B}(B^{+}\to \jpsi K^{+})$ and ${\cal
  B}(\jpsi\to p\bar p)$ branching
fractions~\cite{Beringer:1900zz}. For the charmonium modes we compare
our values to the product of the indipendently measured branching
fractions. The first uncertainties are statistical,
the second systematic in the present measurement, and the third systematic from the
uncertainty on the $J/\psi$ branching fraction.}
\renewcommand{\arraystretch}{1.1}
\vskip 0.3cm
\centering
\begin{tabular}{l|r@{$\pm$}r@{$\pm$}r@{$\pm$}r|c|c}
$B^{+}$ &  \multicolumn{4}{c|}{${\cal B}(B^{+}\to ({\rm mode})\to p\bar p K^{+})$}   & UL $(95\%$ CL) & Previous measurements\\
decay mode &  \multicolumn{4}{c|}{ ($\times 10^6$)} & ($\times 10^6$) & ($\times 10^6$)~\cite{Aubert:2005gw, Wei:2007fg}\\
\hline
total &
$10.81\,$&$\,0.42\,$&$\,0.30\,$&$\,0.49$ & & $10.76^{+0.36}_{-0.33} \pm 0.70$ 
\\
{\scriptsize ${\mpp \unitm{GeV/}c^{2}}$}  &
$4.46 \,$&$\, 0.21\,$&$\,
0.18 \,$&$\, 0.20$ & & $5.12 \pm 0.31$
\\ 
$\eta_{c}(1S) K^{+}$ &
$1.27 \,$&$\, 0.08 \,$&$\,
0.05 \,$&$\, 0.06\,$  & & $1.54 \pm 0.16$
\\ 
$\psi(2S)K^{+}$ &
$0.175\,$&$\, 0.027\, $&$\, 0.020 \,$&$\, 0.008$ & & $0.176 \pm 0.012$
\\ 
$\eta_{c}(2S)K^{+}$  &
$0.063 \,$&$\,0.025\, $&$\, 0.009\, $&$\, 0.003$
& $<0.106$ \\
$\chi_{c0}(1P)K^{+}$ &
$0.024 \,$&$ \,0.021\,$&$\,
0.006 \,$&$\, 0.001$
& $<0.062$ & $0.030 \pm 0.004$ \\ 
$h_{c}(1P)K^{+}$ &
$0.034 \,$&$\, 0.018\,$&$\, 0.008\,$&$\, 0.002$ 
& $<0.064$\\ 
$X(3872)K^{+}$ &
$-0.015\, $&$\, 0.013\,$&$\,
0.003 \,$&$ \,0.001$ 
& $<0.017$\\ 
$X(3915)K^{+}$ &
$0.022 \,$&$ \,0.029\, $&$\, 0.004\, $&$\, 0.001$ 
& $<0.071$\\ 
\end{tabular}
\label{tab:2011}
\end{table}

\noindent The branching fractions obtained are compatible with the world average
values~\cite{Beringer:1900zz}. The upper limit on 
${\mathcal B}(B^{+} \to \chi_{c0}(1P) K^{+} \to p \bar p K^{+})$ is
compatible with the world average ${\mathcal B}(B^{+} \to \chi_{c0}(1P) K^{+}) \times {\mathcal
B}(\chi_{c0}(1P) \to p \bar p) = (0.030 \pm 0.004) \times 10^{-6}$~\cite{Beringer:1900zz}.
We combine our upper limit for $X(3872)$ with the known value for ${\mathcal B} (B^{+} \to X(3872) K^{+} ) \times {\mathcal B} (X(3872) \to
J/\psi \pi^{+} \pi^{-})= (8.6 \pm 0.8) \times 10^{-6}$~\cite{Beringer:1900zz}
to obtain the limit
\begin{equation*}
\frac{{\mathcal B} (X(3872) \to p \bar p)}{{\mathcal B} (X(3872) \to J/\psi \pi^{+}
  \pi^{-})}< 2.0\times 10^{-3}.
\end{equation*}
This limit challenges some of the
predictions for the molecular interpretations of the $X(3872)$ state
and is approaching the
range of predictions for a conventional $\chi_{c1}(2P)$
state~\cite{Chen:2008cg, Braaten:2007sh}.
Using our result and the $\eta_c(2S)$ branching fraction
${\mathcal B} (B^{+} \to \eta_{c}(2S) K^{+})\times
{\mathcal B} (\eta_{c}(2S) \to K \bar K \pi) = (3.4\, ^{+2.3}_{-1.6}) \times 10^{-6}$~\cite{Beringer:1900zz},
a limit of
\begin{equation*} 
\frac{{\mathcal B} (\eta_{c}(2S)
\to p \bar p)}{{\mathcal B} (\eta_{c}(2S) \to K \bar K \pi)} < 3.1 \times 10^{-2}
\end{equation*}
is obtained.

\section{Summary}
Based on a sample of $6951 \pm 176$ $B^{+} \to p \bar p K^{+}$ decays
reconstructed in a data sample, corresponding to an integrated
luminosity of
$1.0\unitm{fb^{-1}}$, collected with the LHCb detector, the following
relative branching fractions are measured
\begin{align*}
\frac{{\mathcal B}(\bppk)_{\rm total}}{{\mathcal B}(B^{+} \to J/\psi K^{+} \to p
\bar p K^{+})}=& \, 4.91 \pm 0.19 \, {(\rm
  stat)} \pm 0.14 \,{(\rm syst)},\\
\frac{{\mathcal B}(\bppk)_{\mpp \unitm{GeV/}c^{2}}}{{\mathcal B}(B^{+} \to J/\psi K^{+} \to p
\bar p K^{+})}=& \, 2.02 \pm 0.10 \,{(\rm
  stat)}\pm 0.08 \, {(\rm syst)},\\
\frac{{\mathcal B} (B^{+} \to \eta_{c}(1S) K^{+} \to p
\bar p K^{+})}{{\mathcal B}(B^{+} \to J/\psi K^{+} \to p
\bar p K^{+})} = & \, 0.578 \pm 0.035 \, {(\rm stat)} \pm 0.025 \, {(\rm syst)},\\
\frac{{\mathcal B} (B^{+} \to \psi(2S)
K^{+} \to p
\bar p K^{+})}{{\mathcal B}(B^{+} \to J/\psi K^{+} \to p
\bar p K^{+})}=& \, 0.080 \pm 0.012 \, {(\rm stat)} \pm 0.009 \, {(\rm syst)}.\\
\end{align*}
An upper limit on the ratio $\frac{{\mathcal B} (B^{+} \to X(3872)
K^{+} \to p
\bar p K^{+})}{{\mathcal B}(B^{+} \to J/\psi K^{+} \to p
\bar p K^{+})} < 0.017$ is obtained, from which a
limit of
\begin{equation*}
\frac{{\mathcal B} (X(3872) \to p \bar p)}{{\mathcal B} (X(3872) \to J/\psi \pi^{+}
  \pi^{-})}< 2.0\times 10^{-3}
\end{equation*}
is derived.
\section*{Acknowledgements}

\noindent We express our gratitude to our colleagues in the CERN
accelerator departments for the excellent performance of the LHC. We
thank the technical and administrative staff at the LHCb
institutes. We acknowledge support from CERN and from the national
agencies: CAPES, CNPq, FAPERJ and FINEP (Brazil); NSFC (China);
CNRS/IN2P3 and Region Auvergne (France); BMBF, DFG, HGF and MPG
(Germany); SFI (Ireland); INFN (Italy); FOM and NWO (The Netherlands);
SCSR (Poland); ANCS/IFA (Romania); MinES, Rosatom, RFBR and NRC
``Kurchatov Institute'' (Russia); MinECo, XuntaGal and GENCAT (Spain);
SNSF and SER (Switzerland); NAS Ukraine (Ukraine); STFC (United
Kingdom); NSF (USA). We also acknowledge the support received from the
ERC under FP7. The Tier1 computing centres are supported by IN2P3
(France), KIT and BMBF (Germany), INFN (Italy), NWO and SURF (The
Netherlands), PIC (Spain), GridPP (United Kingdom). We are thankful
for the computing resources put at our disposal by Yandex LLC
(Russia), as well as to the communities behind the multiple open
source software packages that we depend on.

\addcontentsline{toc}{section}{References}
\bibliographystyle{LHCb}
\bibliography{ref}

\end{document}